\documentclass[aps,prl,twocolumn,amsmath,amssymb]{revtex4}
\usepackage{graphicx}
\usepackage{amsmath}
\usepackage{color}
\usepackage[normalem]{ulem}
\def\cp#1{\mathbf{#1}}
\begin{document}

\title{Spin-Orbit Coupled Ultracold Gases in Optical Lattices: \\ High-Band Physics and Insufficiency of Tight-Binding Models} 
\author{Lihong Zhou and Xiaoling Cui}
\email{xlcui@iphy.ac.cn}
\affiliation{Beijing National Laboratory for Condensed Matter Physics, Institute of Physics, Chinese Academy of Sciences, Beijing, 100190, People's Republic of China}

\date{\today}
\begin{abstract}
We study the interplay effect of spin-orbit coupling(SOC) and optical lattice to the single-particle physics and superfluid-insulator transition in ultracold Fermi gases. We consider the type of SOC that has been realized in cold atoms experiments via two-photon Raman processes. Our analyses are based on the knowledge of full single-particle spectrum in lattices, without relying on any tight-binding approximation.
We evaluate existing tight-binding models and point out their limitations in predicting the correct single-particle physics due to the missed high-band contributions.
Moreover, we show that the Raman field (creating SOC) can induce band-gap closing in a two-dimensional optical lattice, leading to the intriguing phenomenon of superfluidity-reentrance for interacting fermions at integer filling. 
We present the superfluid-insulator phase diagram in a wide parameter regime of chemical potentials and Raman fields. All these results are far beyond any tight-binding model can predict, and can be directly probed in current cold atoms experiments.
\end{abstract}
\maketitle

As two typical potentials engineered in cold atoms, the spin-orbit coupling(SOC) and optical lattice both significantly modify the single-particle dispersion and give rise to intriguing collective phenomena in interacting many-body systems\cite{OL_review, SOC_review}. Their combination, i.e., spin-orbit coupled quantum gases in optical lattices, have recently attracted considerable attention in view of their experimental realizations through laser-assistant tunneling\cite{Bloch_1, Bloch_2, Ketterle_1, Ketterle_2}, shaken optical lattices\cite{Sengstock_1, Sengstock_2}, and two-photon Raman processes\cite{edge1,edge2, Engels}. A fascinating property of such system is that the effective flux in each plaquette is on the order of one flux quanta, large enough to reach the quantum Hall regime and enable the exploration of topological signatures\cite{Bloch_1, Bloch_2, Ketterle_1, Ketterle_2, Sengstock_1, Sengstock_2,edge1,edge2}.
Theoretical studies have also revealed various interaction effects in such system\cite{model_i_1, model_i_6, model_i_3, model_i_4, model_i_5, model_i_7, model_i_8, model_i_9, model_i_10, model_ii_2, model_ii_3, model_ii_4, model_ii_5},
and in particular, pointed out the possibilities of majorana fermions\cite{model_i_10, model_ii_5}  and exotic spin textures\cite{model_i_5, model_i_8}.

To date, most studies on the spin-orbit coupled atomic gases in optical lattices have concentrated on the lowest-band physics under various tight-binding approximations\cite{model_i_1, model_i_6, model_i_3, model_i_4, model_i_5, model_i_7, model_i_8, model_i_9, model_i_10, model_ii_2, model_ii_3, model_ii_4, model_ii_5}, while the high-band physics has been rarely explored. Moreover, even for the lowest band(s), it is questionable whether the conventional Wannier wavefunction without SOC can still be used to construct the tight-binding model, as the SOC can induce coupling between different bands and the original Wannier basis could be problematic. With these motivations, in this work we will go beyond the tight-binding approximation and lowest-band physics to explore the interplay effects of SOC and optical lattice. Our study will be based on the knowledge of full single-particle spectrum, and thus will take into account all high-band contributions missed in previous studies. When come to the lowest band(s), this treatment also allows us to test the validity of tight-binding models used in literature\cite{model_i_1, model_i_6, model_i_3, model_i_4, model_i_5, model_i_7, model_i_8, model_i_9, model_i_10, model_ii_2, model_ii_3, model_ii_4, model_ii_5}.

We consider the 
type of SOC that has been realized via two-photon Raman processes in optical lattices\cite{edge1,edge2, Engels}.
We exactly solve the single-particle spectrum, and point out that the existing tight-binding models 
have various limitations in predicting correct single-particle physics. This can be attributed to the SOC-induced excitations to higher bands, or equivalently, the improper usage of Wannier basis in the presence of SOC.
In addition, we find a remarkable feature in the high-band physics, i.e., the Raman-induced gap closing in a two-dimensional(2D) optical lattice.
With attractive interaction between two-species fermions, the gap closing leads to the reentrance of superfluidity from insulating phase at integer filling (two atoms per site). 
We identify the superfluid-insulator phase boundaries in a wide parameter regime of chemical potential and Raman field, and show rich density distributions in the trapped system as varying Raman fields. These results reveal the intriguing single-particle and many-body physics due to the interplay between SOC and optical lattice, which are far beyond any tight-binding model can predict. Our results can be directly probed in current cold atom experiments.

{\it Single-particle physics.} We will first address the single-particle physics in the absence of interaction. The spin-orbit coupled atoms in a one-dimensional(1D) optical lattice can be described by the Hamiltonian
(set $\hbar=1$)
\begin{align}
H_x&=\frac{1}{2m}(p_x-q\sigma_z)^2+\Omega_R\sigma_x+V_0\cos^2(k_L x),
\end{align}
here $q$ is the momentum transferred through the Raman processes
creating SOC, and $\Omega_R$ is the Raman field strength; $V_0$ is
the lattice depth; $k_L$ is the recoil momentum giving the lattice
spacing $a=\pi/k_L$ and recoil energy $E_L=k_L^2/(2m)$.
It is straightforward to exactly solve the eigen-system of Eq.(1) by
writing the field operator as $\psi_{\sigma}(x)=\sum_{nk}
\phi^{\sigma}_{nk}(x)\psi_{nk}$, with the Bloch wave function
$\phi^{\sigma}_{nk}(x)=\frac{1}{\sqrt{L}}\sum_{G} a^{\sigma}_{nk}(G)
e^{i(k+G)x}$, where $n$ is the band index, $k$ is the crystal
momentum $(\in[-k_L,k_L])$, and $G$ is the reciprocal vector. In
each $\{n,k\}$ sector, the Hamiltonian can be diagonalized in the
$\{G,\sigma\}$ subspace\cite{supple}.
In this work we consider the interesting case of $k_L=2q$ , where an infinitesimal $\Omega_R$ can generate a gap between the lowest two bands (see Fig.1).

To capture the low-energy physics in deep lattices,  a traditional way is to expand the field operator by the Wannier function of the lowest band,  $\psi_{\sigma}(x)=\sum_{i} \omega_{n=0,\sigma}(x-x_i) c_{i\sigma}$ ($i$ is lattice site), so to result in certain tight-binding (lattice) model.  For the present case, two types of such models
have been employed
 \cite{model_i_1, model_i_6, model_i_3, model_i_4, model_i_5, model_i_7, model_i_8, model_i_9, model_i_10, model_ii_2, model_ii_3, model_ii_4, model_ii_5}:
\begin{eqnarray}
(A) \ H&=&-t\sum_i(e^{-iqa}c_{i,\uparrow}^{\dag}c_{i+1,\uparrow}+e^{iqa}c_{i,\downarrow}^{\dag}c_{i+1,\downarrow}+h.c.)\nonumber\\
&&+\Omega\sum_i (c_{i,\uparrow}^{\dag}c_{i,\downarrow}+h.c.); \label{H1}\\
(B) \ H&=&-t\sum_{i\sigma}(c_{i,\sigma}^{\dag}c_{i+1,\sigma}+h.c.)+\Omega_R\sum_i (c_{i,\uparrow}^{\dag}c_{i,\downarrow}+h.c.)\nonumber\\
&&+i t' \sum_i \left[(c_{i,\uparrow}^{\dag}c_{i+1,\uparrow} - c_{i,\downarrow}^{\dag}c_{i+1,\downarrow})-h.c.\right]. \label{H2}
\end{eqnarray}
Model (A) 
uses a shifted Wannier basis $\omega_{n\sigma}(x)=e^{i\nu_{\sigma} qx}\omega_n(x)$ ($\nu_{\updownarrow}=\pm1,\ n=0$), i.e., the conventional Wannier function with a momentum shift for each spin;
$t$ is the hopping amplitude identical to that without SOC\cite{OL_review}; $\Omega=\Omega_R \int dx e^{i2qx}|\omega_0(x)|^2$ is the on-site spin-flip. 
This is equivalent to the idea of Peierls substitution under the gauge potential $A_x=q\sigma_z$\cite{Goldman}. In comparison, Model (B) 
uses the conventional Wannier basis $\omega_{n=0}(x)$, with $t$ the same as in (A) and $t'=\frac{q}{m} \int dx \omega_0(x) \frac{\partial}{\partial x}\omega_0(x-a)$\cite{model_ii_2}. Model (A) and (B) are fundamentally different as they are based on distinct Wannier bases\cite{supple}.

\begin{figure}[t]
\includegraphics[width=9.5cm]{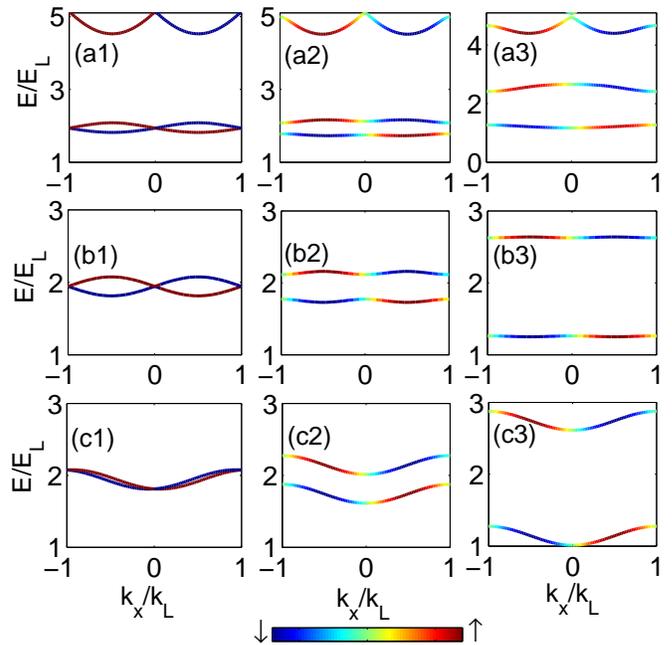}
\caption{(Color online) Single-particle energy spectrum in a 1D optical lattice with $V_0=5E_L$, based on the exact Hamiltonian (first panel) and the tight-binding models (A) and (B) (second and third panels). The model plots have incorporated the constant shifts when reducing the exact Hamiltonian to the tight-binding ones\cite{supple}. The first, second, and third columns are respectively with $\Omega_R/E_L=0,\ 0.2,\ 0.8$.
 The color of the curves represents the weight of different spins in the corresponding wave function (red for $\uparrow$ and blue for $\downarrow$).}  \label{fig1}
\end{figure}

In Fig.1, we compare the single-particle dispersion exactly solved from Eq.(1) (see Fig.1(a1-a3)) with those from tight-binding models (A) and (B) ((b1-b3) and (c1-c3)) for several values of $\Omega_R$. At $\Omega_R=0$, Fig.1(a1) shows that 
the spectra of up- and down-spin are respectively shifted right- and left-side by a momentum $q=k_L/2$, compared to that without SOC, and they cross each other at $k=0,\pm k_L$. These shifts are due to the fact that $p_x\sigma_z$ term in Eq. (1) can be gauged away by a unitary transformation $\hat{U}=e^{iqx\sigma_z}$.
With an infinitesimal $\Omega_R$,
energy gaps will open at these crossing places,
see Fig.1(a2). These features can be well captured by model (A) using the shifted Wannier basis, as shown by Fig.1(b1,b2).
When increasing $\Omega_R$ further, however, model (A) fails to reproduce the correct spectrum. At large $\Omega_R$, the spin tends to be polarized and the spectrum approaches to that of a spinless particle in a lattice potential, with the lowest two bands respectively featuring s- and p-orbitals (as indicated by Fig.1(a3)). On the contrary, 
model (A) always produces similar band structure as in small $\Omega_R$ case, merely with narrower band widths and larger band gap (Fig.1(b3)).
For model (B), we see from Fig.1(c1-c3) that it cannot reproduce the correct single-particle physics for any value of $\Omega_R$.  At zero $\Omega_R$, the momentum shift, as given by $t'$ ($\ll t$\cite{supple}), is hardly visible in Fig.1(c1); moreover, $t'$ cannot be fully gauged away here, in contrast to the continuum case and model (A). Thus, model (B) is inapplicable when SOC is present.

The breakdown of model (A) at finite $\Omega_R$, as manifest by obvious discrepancies between Fig.1(a3) and (b3), is naturally related to the higher-band contributions missed in this model. 
This can be seen clearly in the large $\Omega_R$ limit, where the correct Wannier wavefunction to approximate the lowest band(s) should be with the spin transversely polarized and the orbit identical to the conventional $\omega_0(x)$.
Importantly, this wavefunction is a superposition of all levels of shifted Wannier bases $\{\omega_{n\sigma}(x)\}$, which means that many higher bands will contribute to the low-energy physics. 
Similarly, the breakdown of model (B) can also be analyzed. 
For instance, at zero $\Omega_R$, the $p_x\sigma_z$ term can induce excitations to higher bands. The resulted spectra turn out to be well described by the shifted Wannier basis, which is a superposition of all levels of conventional ones.

We remark that the insufficiency or inapplicability of various tight-binding models is a direct consequence of the interplay between SOC and optical lattice. SOC couples the spin and orbital degrees of freedom, while in a lattice setup the spin flip is directly associated with the band excitation. If the SOC-induced band excitations are very serious, the tight-binding model based on the original Wannier basis will no longer be valid. This mechanism is expected to hold for a general type of SOC and background setup\cite{Yi}.

\begin{figure}[h]
\centering
\includegraphics[width=9.5cm]{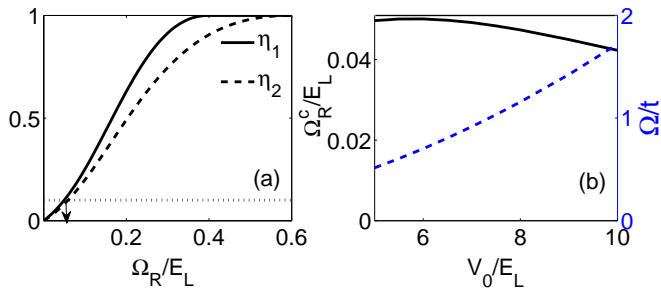}
\caption{(Color online) Validity of tight-binding model (A). (a) $\eta_{\nu}\ (\nu=1,2)$ as a function of $\Omega_R$ for given  $V_0=5E_L$. The arrow locates the critical Raman field $\Omega^c_R$ (see text).  (b) $\Omega^c_R$ and the corresponding ratio $\Omega/t$ as a function of $V_0$.  } \label{fig2}
\end{figure}

It is useful to identify a quantitative parameter regime for model (A) to be valid. Intuitively, one would compare $\Omega_R$ with the width of lowest band(s) and the gap to higher bands. Nevertheless, a large gap is generally associated with a narrow width, thus it is not immediately clear how the requirement of $\Omega_R$ changes with $V_0$. We thus carry out a numerical analysis to the band structure and then compare with model predictions.
Based on the dispersion of model (A) \cite{supple}, $k=0$ and $k=\pm k_L$ always have the same energy and each band has double maxima and double minima (Fig.1(b2,b3)). However, exact solutions show that as $\Omega_R$ increases, the energies at $k=0$ and $k=\pm k_L$ become more and more deviated, and finally the double minima vanish in the first(lowest) band and double maxima vanish in the second  band (Fig.1(a3)). To describe this feature, we introduce two quantities
\begin {eqnarray}
\eta_{\nu}&=&\frac{|E_{\nu}(k=0)-E_{\nu}(k=\pm k_L)|}{W_{\nu}}\ \ \ \ (\nu=1,2) ,
\end{eqnarray}
here $W_{\nu}$ is the band width from exact solutions. $\eta_{\nu}$ is always zero according to model (A), while is generally finite according to exact spectrum.

In Fig.2a, we show that at given $V_0$,
both $\eta_1$ and $\eta_2$ increase with $\Omega_R$ and finally saturate at unity. We determine a critical Raman field ($\Omega_R^c$) for the validity of model (A) by requiring both $\eta_1$ and $\eta_2$ below $10\%$. In Fig.2b, we plot $\Omega_R^c$ as a function of $V_0$ and its corresponding $(\Omega/t)_c$.
We can see that for $V_0\in(5,10)E_L$, $\Omega_R^c$ varies in a narrow region between $0.04$ and $0.05E_L$, comparable to the lowest-band width ($\in(0.02,0.09)E_L$) but much smaller than the gap to higher bands ($\in(2.43,4.55)E_L$).
Remarkably, such weak SOC strength can already produce considerable deviation of the band structure from the exact solutions.
With such $V_0$ and $\Omega_R^c$, we find $(\Omega/t)_c$ can range from $0.5$ to $1.8$, which sets the upper limit of $\Omega/t$ for model (A) to be valid.

{\it Band evolution in 2D.} The evolution of 1D spectrum as tuning $\Omega_R$ (shown in Fig.1(a1-c1)) can induce even interesting band evolution in higher-D optical lattices, such as in 2D lattices when adding another lattice potential along $y-$direction, say, $V(y)=V_0 \cos^2(k_L y)$.
Generally, the lowest band in x-direction and the lowest band in y-direction, noted as $1x+1y$, comprise the lowest band in 2D; using similar notations, $2x+1y$ and $1x+2y$ respectively comprise the second and third lowest bands, see Fig.3(a1)(a2). As increasing $\Omega_R$, the gap between $1x$ and $2x$ increases, leading to non-trivial band-gap evolutions.
As shown in Fig.3b, the first band gap opens at $\Omega_R=0.175E_L$ when the maximum energy of the first band (at $(k_x,k_y)=(\pm k_L,\pm k_L$)) matches the minimum of the second band (at $(\pm k_L,0)$), while the second band gap closes at $\Omega_R=1.35E_L$ when the second and the third band touch at $(0,\pm k_L)$.
Note that the tight-binding model (A) tend to overestimate the first gap as increasing $\Omega$ (see Fig.3b), and meanwhile, cannot predict the closing of the second band gap. Particularly, the band-gap closing reflects the unique high-band physics for the spin-orbit coupled atoms in high-D lattices, which turns out to induce intriguing many-body physics as shown below.

\begin{figure}[t]
\includegraphics[width=8cm]{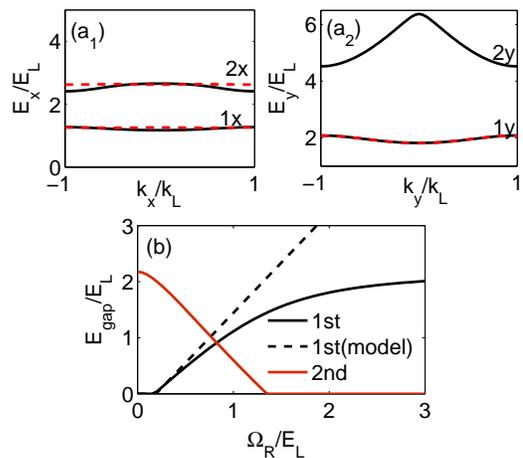}
\caption{(Color online) Band-gap evolution in $2D$ optical lattices. The energy dispersion can be decomposed as $E_{\bf k}=E_x(k_x)+E_y(k_y)$. (a1,a2) $E_x$ and $E_y$ for the lowest two bands at $V_0=5E_L$ and $\Omega_R=0.8E_L$. The band index is marked accordingly (see text). Red dashed lines show the prediction of model (A) to the band $1x,\ 2x$ and $1y$. (b) The first (between $1x+1y$ and $2x+1y$) and second (between $2x+1y$ and $1x+2y$) band gap as functions of $\Omega_R$ at $V_0=5E_L$.  Dashed line shows the prediction of model (A) to the first band gap. }\label{fig3.eps}
\end{figure}


{\it Superfluid-Insulator transition.} 
We investigate the superfluidity of fermions loaded into a 2D optical lattice. The interacting Hamiltonian is
\begin{equation}
H=\sum_{\sigma\sigma'}\int d\cp{r}\psi_{\sigma}^\dagger(\cp r)H_0(\cp r)\psi_{\sigma'}(\cp r)+g\int d\cp{r}\psi_{\uparrow}^\dagger(\cp r)\psi_{\downarrow}^\dagger(\cp r)\psi_{\downarrow}(\cp r)\psi_{\uparrow}(\cp r);
\end{equation}
Here $H_0({\cp r})=H_x+H_y$, with ${\cp r}=(x,y)$ and $H_y=p_y^2/(2m)+V_0\cos^2(k_Ly)$; the bare interaction $g$ can be related to the 2D binding energy $E_b$ by the renormalization equation $g^{-1}=-S^{-1}\sum_{\cp k} ({\cp k}^2/m+E_b)^{-1}$ ($S$ is the quantization area).

Following the treatment of fermion superfluidity in optical lattices\cite{Zhai, Moon, Cui}, we consider the dominated intra-band pairing with the opposite crystal momenta and the same energy, which gives the pairing amplitudes
\begin{equation}
\Delta_{\cp G}=-\frac{g}{S} \sum_{\cp{nk}} M^{\cp G}_{\cp{nk}} \langle \psi_{\cp{n -k}} \psi_{\cp{n k}} \rangle;\ \ \Delta_{\cp{nk}}=\sum_{\cp G} \Delta_{\cp G} M^{{\cp G}*}_{\cp{nk}}. \label{Delta}
\end{equation}
with $M^{\cp G}_{\cp{nk}}=\sum_{\cp Q} a^{\downarrow}_{\cp{n-k}}(-{\cp Q}) a^{\uparrow}_{\cp{nk}}(\cp{Q+G})$. Here $\psi_{\cp{n k}}$ and $a^{\sigma}_{\cp{nk}}$ are the 2D analogs of $\psi_{n k}$ and $a^{\sigma}_{nk}$ as specified in the 1D case.
Given Eq.(\ref{Delta}), the thermodynamic potential ${\cal K}=\langle H-\mu N\rangle$ can be calculated as
\begin{eqnarray}
\cal{K}&=&\sum_{\cp{n}, k_x>0, k_y} \left( E_{\cp{nk}} -\mu -\sqrt{(E_{\cp{nk}} -\mu)^2+|\Delta_{\cp{nk}}-\Delta_{\cp{n-k}}|^2}  \right)  \nonumber \\
&& - \frac{S}{g} \sum_{\cp G} |\Delta_{\cp G}|^2, \label{K}
\end{eqnarray}
where $E_{\cp{nk}}$ is the single-particle eigen-energy and $\mu$ is the chemical potential. The ground state can be obtained by minimizing $\cal{K}$ with respect to $\{\Delta_{\cp G}\}$. In practice we have kept the lowest three ${\cp G}=(0,0),\ (2k_L,0),\ (0,2k_L)$. We have checked that the results obtained will not be visibly altered by including more $\Delta_{\cp G}$ into the minimization.


\begin{figure}[t]
\includegraphics[width=8cm]{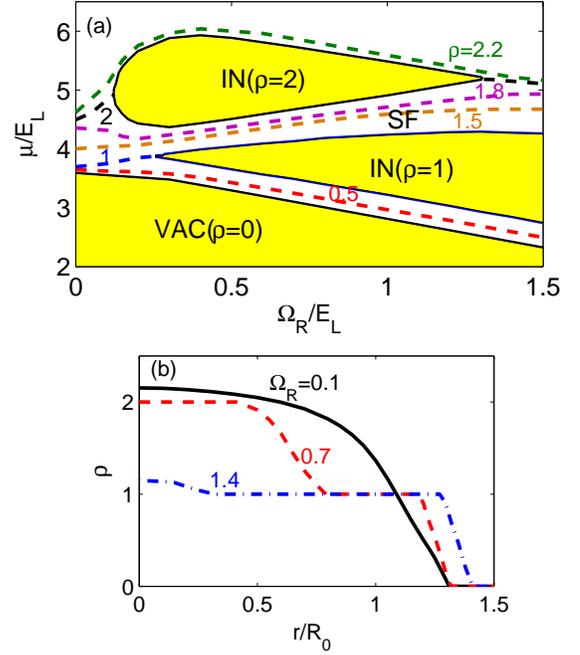}
\caption{(Color online) (a)Superfluid-insulator phase diagram
with $V_0=5E_L$ and $E_b=3E_L$.  Contour-density plots are shown for filling factors $\rho=2.2,2,1.8,1.5,1,0.5$ (top to bottom). 
(b) Density profiles of $^{40}$K atoms in a trapped system at $\Omega_R=0.1,0.7,1.4E_L$. $V_0$ and $E_b$ are the same as in Fig.4(a). We consider the trapping frequency $\omega=(2\pi)140$HZ and the lattice spacing $a=425$nm, giving $R_0=\sqrt{2E_L/(m\omega^2)}=100/k_L$. The total particle number is fixed as $5\times 10^4$. }\label{fig5}
\end{figure}

Near the Superfluid(SF)-Insulator(IN) transition, all $\{\Delta_{\cp G}\}$ continuously evolve to zero, thus one can expand $\cal{K}$ as $\cal{K}$$=\sum_{\cp{GG'}}\Delta_{\cp G}^*C_{\cp{GG'}}\Delta_{\cp G'}$, with the matrix
\begin{equation}
C_{\cp{GG'}}=-\frac{S}{g}\delta_{\cp{GG'}}-\sum_{\cp{n},k_x>0,k_y}\frac{(M_{\cp{nk}}^{\cp{G}*}
-M_{\cp{n-k}}^{\cp{G}*})(M_{\cp{nk}}^{\cp{G'}}-M_{\cp{n-k}}^{\cp{G'}})}{2|E_{\cp{nk}}-\mu|}
\end{equation}
The SF-IN transition can thus be determined by setting the determinant of $C-$matrix to be zero.

In Fig.4a, we present the SF-IN phase diagram in the $\mu-\Omega_R$ parameter plane with given $V_0=5E_L$ and $E_b=3E_L$. A remarkable feature is that the IN phase with filling $\rho=2$ appears as an isolated area surrounded by the SF bath in the phase diagram. This suggests that a system with fixed $\rho=2$ will undergo a sequence of transitions from SF to IN and to SF again when increasing $\Omega_R$, as explicitly shown by the density contour plot in Fig.4a. This can be attributed to two competitive effects induced by $\Omega_R$:
the spin-polarization does not favor the s-wave pairing and thus gives rise to the first transition, while the vanishing band gap (as shown in Fig.3b) induces pairing around the Fermi surface and gives the second transition. Here, we can see that the single-particle physics  manifest itself well in the SF-reentrance of interacting many-body systems.

Fig.4a also shows the SF to IN transition at half-filling ($\rho=1$) as changing $\Omega_R$. This is due to its two cooperative effects: the spin polarization and the increasing (first) band gap, both of which disfavor the pairing superfluid. Though the model (A) can also capture these two effects, we do not expect it give the correct SF-IN phase boundary at $\Omega_R\gtrsim E_L$, as the band gap is considerably overestimated in that regime (see Fig.3b).

The $(\mu, \Omega_R)$ diagram implies very rich density profiles in a trapped system as varying $\Omega_R$. In Fig.4b, we show several typical density profiles for trapped $^{40}$K atoms, obtained by applying the local density approximation with  $\mu(\cp{r})=\mu(0)-V(\cp{r})$ at position $\cp{r}$, and $V(\cp{r})=m\omega^2 \cp{r}^2/2$ is the external harmonic trap.
We can see that as increasing $\Omega_R$, the $\rho=2$ and $\rho=1$ IN phases (with flat-top density) respectively emerge near the center and edge of the trap. Increasing $\Omega_R$ further, the $\rho=2$ IN phase gives way to SF and the flat-top structure disappears. This is directly related to the gap-closing and SF-reentrance discussed above. Eventually, at large $\Omega_R$ the $\rho=1$ IN phase will occupy a large region in the trap.

{\it Summary.} In summary, we have studied the single-particle physics and the superfluid-insulator transition due to the interplay of SOC and optical lattices. We point out various limitations of existing tight-binding models in predicting the correct single-particle physics. We also reveal interesting high-band physics that have been missed in previous studies, including the SOC-driven band-gap evolution and the resulted intriguing SF-IN transitions, which are far beyond any tight-binding model can predict. These results can be directly explored in current cold atom experiments. For instance, the single-particle spectrum can be probed by the momentum-resolved Bragg spectroscopy\cite{Sengstock};
the SF-IN transition and the reentrance physics can be detected by measuring the momentum space condensation fraction similar to the continuum case\cite{Jin1}, or can be inferred from the density profile in a trapped system.


{\it Acknowledgements.}
We acknowledge support from NSFC 11374177 and the programs of Chinese Academy of Sciences.

\end{document}


\title{Supplementary Materials}
\author{}
\affiliation{}
\maketitle

In this supplementary material we provide more details about how to exactly solve the single-particle Hamiltonian and to derive the two tight-binding models.

\subsection{I. Exactly solving the single-particle Hamiltonian}

We first write down the second-quantized Hamiltonian for $H_x$ (Eq.(1) in the main text):
\begin{align}
H=\int dx\sum_{\sigma=\uparrow,\downarrow}\psi_{\sigma}^\dagger( x)\hat{H}_x(x)\psi_{\sigma}(x)
\end{align}
We can expand the field operator  in terms of Bloch wave functions of $\hat{H}_x$, $\psi_\sigma(x)=\sum_{nk}\phi_{nk}^\sigma(x)\psi_{n\cp k}$, with k lying within the first Brillouin zone (BZ) and n the band index. According to the Bloch theorem, one can further expand the Bloch wave function in terms of plane waves, $\phi_{n k }^\sigma(x)=\frac{1}{\sqrt{L}}\sum_{G}a_{n k}^\sigma(G)e^{i (k+G)x}$, with $G$ the reciprocal vector (integer of $2k_L$). Then the Hamiltonian $H$ can be written as
 \begin{align}
H&=\sum_{n,k\in BZ}\sum_{ G'}a_{nk}^{\uparrow\ast}(G')\sum_{ G}a_{n k}^\uparrow( G)\{[\frac{\hbar^2}{2m}(k+G-q)^2+\frac{1}{2}V_0]\delta_{{GG'}}+\frac{V_0}{4}\delta_{ G\pm2k_L, G'}\}\psi_{nk}^\dagger\psi_{n k}\nonumber\\
&+\sum_{n k\in BZ}\sum_{ G'}a_{n k}^{\downarrow\ast}(G')\sum_{ G}a_{n k}^\downarrow( G)\{[\frac{\hbar^2}{2m}({k+G}+q)^2+\frac{1}{2}V_0]\delta_{{GG'}}+\frac{V_0}{4}\delta_{ G\pm2k_L,G'}\}\psi_{n k}^\dagger\psi_{n k}\nonumber\\
&+\sum_{n,k\in BZ}[\sum_{ G'}a_{nk}^{\uparrow\ast}(G')\sum_{ G}a_{n k}^\downarrow( G)\delta_{GG'}\Omega_R\psi_{nk}^\dagger\psi_{n\cp k}+\sum_{ G'}a_{nk}^{\downarrow\ast}(G')\sum_{ G}a_{n k}^\uparrow( G)\delta_{GG'}\Omega_R\psi_{nk}^\dagger\psi_{nk}]
\end{align}
In each $\{n,k\}$ sector, $H_{nk}=\sum_{GG'}\sum_{\sigma\sigma'}a_{nk}^{\sigma\ast}(G)a_{nk}^{\sigma'}(G')H_{GG'}^{\sigma\sigma'}$, the $H_{GG'}^{\sigma\sigma'}$ is
\begin{equation}
 H_{GG'}^{\sigma\sigma'}=\left(
  \begin{array}{ccccccccc}
    \varepsilon_{k-N2k_L}^\uparrow&{V_0}/{4}&0&...&0&\Omega_R&0&\dots&0\\
   V_0/4& \varepsilon_{k-(N-1)2k_L}^\uparrow&V_0/4&\dots&0&0&\Omega_R&\dots&0\\
    \vdots&\vdots&\vdots&\dots&\vdots&\vdots&\vdots&\vdots&\vdots\\
    0&0&0&\dots&\varepsilon_{k+N2k_L}^\uparrow&0&0&...&\Omega_R\\
    \Omega_R&0&0&\dots&0&\varepsilon_{k-N2k_L}^\downarrow&{V_0}/{4}&\dots&0\\
    0&\Omega_R&0&\dots&0&V_0/4&\varepsilon_{k-(N-1)2k_L}^\downarrow&\dots&0\\
    \vdots&\vdots&\vdots&\dots&\vdots&\vdots&\vdots&\vdots&\vdots\\
    0&0&0&\dots&\Omega_R&0&0&\dots&\varepsilon_{k+N2k_L}^\downarrow
  \end{array}
\right)
\end{equation}
 Here $\varepsilon_k^\updownarrow=\frac{\hbar^2}{2m}(k\mp q)^2+\frac{V_0}{2}$, and $H_{GG'}^{\sigma\sigma'}$ is a $(4N+2)\times(4N+2)$ matrix, where $N$ is a cutoff number and in our practical calculation we take $N=50$. A larger N will not visibly change the spectra for the bands that we are interested in.

\subsection{II. Deriving the tight-binding models (A) and (B)}

To derive the tight-binding model, we expand the field operator in terms of Wannier function of the lowest band, $\psi_{\sigma}(x)=\sum_{i}\omega_{n=0,\sigma}(x-x_i)c_{i\sigma}$, here $i$ is the index of lattice site. Given the expression of the Wannier function $\omega_{n=0,\sigma}(x-x_i)$, the second-quantized form of the Hamiltonian can be reduced to a certain tight-banding model.

  \subsubsection{ Model (A) }

In this model, we take the shifted Wannier basis
\begin{equation}
\omega_{n\sigma}(x-x_i)=e^{i\nu_\sigma q (x-x_i)}\omega_n(x-x_i),
\end{equation}
with $\nu_\updownarrow=\pm1$ and $n=0$. The parameters in the model (A) include:

(1) the nearest-neighbor hopping term:
\begin{align}
t_{i,i+1}^\sigma&=-\int dx \omega_{0\sigma}^*(x-x_i)[\frac{1}{2m}(p_x-q\sigma_z)^2+V_0\cos^2{(k_L x)}]\omega_{0\sigma}(x-x_{i+1})\nonumber\\
&=- e^{-i\nu_\sigma q a}  \int dx  \omega_0(x)(\frac{p_x^2}{2m}+V_0\cos^2(k_0x))\omega_0(x-a) \nonumber\\
&\equiv t e^{-i\nu_\sigma q a}
\end{align}
here $t=-\int dx w_0(x)(\frac{p_x^2}{2m}+V_0\cos^2(k_L x ))w_0(x-a)$ is the hopping amplitude identical to that without SOC. To obtain above equation, we have used the relation:
\begin{equation}
e^{-i\nu_\sigma q x}[\frac{1}{2m}(p_x-q\sigma_z)^2+V_0\cos^2{(k_L x)}]e^{i\nu_\sigma q x}=\frac{p_x^2}{2m}+V_0\cos^2(k_Lx).
\end{equation}
Therefore the phase shift in front of the Wannier basis is to gauge away the $p_x\sigma_z$ term, so that the resulted topping only differs by a phase factor compared to the conventional one (without SOC).

(2) the on-site spin-flip term:
 \begin{align}
 \Omega_{\sigma\bar{\sigma}}&=\int dx\omega_{0\sigma}^\ast(x)\Omega_R\omega_{0\bar{\sigma}}(x)
 =\Omega_R\int dx e^{-i2\nu_\sigma q x}|\omega_{0}(x)|^2\nonumber\\
 &=\Omega_R\int dx \cos(2q x)|\omega_{0}(x)|^2\nonumber\\
 &\equiv \Omega
 \end{align}
 In previous studies, the on-site spin-flip $\Omega$ is generally set to be identical to $\Omega_R$, which is a good approximation only for deep enough lattice with very localized Wannier wave function.

In addition, in writing the model (A) (Eq.2 in the main text), we have omitted the effective chemical potential term $\sum_{i\sigma} \epsilon_i^A c_{i\sigma}^\dagger c_{i\sigma}$, with
 \begin{align}
 \epsilon_i^A&=\int dx \omega_{0\sigma}^*(x-x_i)[\frac{1}{2m}(p_x-q\sigma_z)^2+V_0\cos^2{(k_L x)}]\omega_{0\sigma}(x-x_i)\nonumber\\
 &=\int dx  \omega_0(x)(\frac{p_x^2}{2m}+V_0\cos^2(k_0x))\omega_0(x)
 \end{align}
 In comparing the spectrum between exact solution and model solutions (Fig.1), we have incorporated above energy shift in Fig.1(b1-b3).

For the lattice model (A), its dispersion can be obtained straightforwardly by using the Fourier transform ${c}_{i\sigma}=\frac{1}{\sqrt{N}}\sum_{ k}e^{i kx_i}{c}_{k\sigma}$. Finally, we get the spectrum of the lowest two bands in k-space
\begin{equation}
 E^A_{1,2}(k)=\epsilon_k^A\pm\sqrt{4t^2\sin^2(\pi k/k_L)+\Omega^2}.
 \end{equation}

 \subsubsection{Model (B)}

In this model, we take the conventional Wannier basis which is spin-independent:
\begin{equation}
\omega_{n\sigma}(x-x_i)=\omega_n(x-x_i),
\end{equation}
with $n=0$. The parameters in the model (B) include:

(1) nearest-neighbor hopping term induced by $\frac{p_x^2}{2m}+V_0\cos^2{(k_L x)}$, i.e., the first term of model (B) with amplitude $t$.

(2) nearest-neighbor hopping term induced by $p_x\sigma_z$:
 \begin{align}
t_{i,i+1}^{' \sigma}&=\frac{iq\nu_{\sigma}}{m}\int dx\omega_0^*(x-x_i)\frac{\partial}{\partial x}\omega_0( x-x_{i+1})\nonumber\\
&=\frac{iq\nu_{\sigma}}{m}\int dx\omega_0(x)\frac{\partial}{\partial x}\omega_0( x-a)\nonumber\\
&\equiv i t' \nu_{\sigma};
 \end{align}
 where $t'$ is given in the main text. For the deep lattice $V_0\gg E_L$, we find
 \begin{equation}
 t'=\frac{\pi}{2}E_L (\frac{V_0}{E_L})^{\frac{1}{2}}e^{-\frac{1}{4}\pi^2\sqrt{\frac{V_0}{E_L}}},
 \end{equation}
 which is much smaller than $t$ (see Ref.1 cited in the main text) :
  \begin{equation}
 t=\frac{4}{\sqrt{\pi}}E_L (\frac{V_0}{E_L})^{\frac{3}{4}}e^{-2\sqrt{\frac{V_0}{E_L}}}.
 \end{equation}
This results in an invisible momentum shift for the up and down-spin spectra in Fig.1(c1). This analysis also holds for other values of $q$ that is typically of the order of $k_L$ in realistic experiments (Ref.9,10,11 in the main text).

(3) the on-site spin-flip term:
 \begin{align}
 \Omega_{\sigma\bar{\sigma}}&=\int dx\Omega_R|\omega_{0}(x)|^2\nonumber\\
  &\equiv \Omega_R
 \end{align}

Similar to model (A), in writing the model (B) (Eq.3 in the main text), we have omitted the effective chemical potential term $\sum_{i\sigma} \epsilon_i^B c_{i\sigma}^\dagger c_{i\sigma}$, with
 \begin{align}
 \epsilon_i^B&=\int dx \omega_{0}^*(x-x_i)[\frac{1}{2m}p_x^2+V_0\cos^2{(k_L x)} + \frac{q^2}{2m}]\omega_{0}(x-x_i)\nonumber\\
 &= \epsilon_i^A +  \frac{q^2}{2m}
 \end{align}
 In Fig.1(c1-c3), we have incorporated above energy shift.

For the lattice model (B), its dispersion can also be obtained straightforwardly by using the Fourier transform ${c}_{i\sigma}=\frac{1}{\sqrt{N}}\sum_{ k}e^{i kx_i}{c}_{k\sigma}$, and finally we get the spectrum of the lowest two bands:
\begin{equation}
 E^B_{1,2}(k)=\epsilon_k^B-2t\cos(\pi k/k_L)\pm\sqrt{4t'^2\sin^2(\pi k/k_L)+\Omega_R^2}.
  \end{equation}